\documentstyle[prabib,aps,twoside]{revtex}

\newcommand{\be}{\begin{equation}}
\newcommand{\bea}{\begin{eqnarray}}
\newcommand{\ee}{\end{equation}}
\newcommand{\eea}{\end{eqnarray}}

\begin{document}
\title{Spacetime torsion and parity violation: a gauge invariant formulation} 
\author{Biswarup Mukhopadhyaya\footnote{Electronic address: {\em
biswarup@mri.ernet.in}}${}^{(1)}$,
Soumitra SenGupta\footnote{Electronic address: {\em 
soumitra@juphys.ernet.in}}
${}^{(2)}$,  and Saurabh Sur\footnote{Electronic address: {\em 
saurabh@juphys.ernet.in}}${}^{(3)}$}
\address{{\rm $^{1}$}Harish Chandra Research Institute, Chhatnag
Road, Jhusi, Allahabad 211 019, India \\
{\rm $^{2,3}$}Department of Physics, Jadavpur University,
Calcutta 700 032, India}
\maketitle

\begin{abstract}
The possibility of parity violation through spacetime torsion
has been explored in a scenario containing fields  with
different spins. Taking the 
Kalb-Ramond field as the source of torsion,an explicitly parity violating
$U(1)_{EM}$
gauge invariant theory  has been constructed by extending the
KR field with a Chern-Simons term.
\end{abstract}

\bigskip
Parity violation in gravitation is an important possibility opened up by   
torsion in a curved spacetime. Essentially,
an antisymmetric extension of the affine connection 
destroys the cyclicity of the Riemann-Christoffel tensor, thereby 
enabling one to add a term of the form 
~$\epsilon^{\alpha\beta\mu\nu} R_{\alpha\beta\mu\nu}$~ to the Einstein-Cartan (EC)
action \cite{hoj}. Such a term explicitly violates parity
in the gravity sector \cite{hari}. 
Since torsion is often looked upon as an artifact of matter fields  
with spin \cite{sab}, it is desirable to have a consistent scheme of  
incorporating  parity-violating
effects in the  coupling of particles of different spins
with torsion.
Such a scheme was attempted first in \cite{bmss} by augmenting the
EC connection with a general set of pseudo-tensorial terms constructed out of
the torsion tensor itself.  

Though the procedure in \cite{bmss} is straightforward for the pure gravity
sector and for spin-1/2 particles, it is beset with the well-known problem of 
gauge-invariance  \cite{ham}  when one takes up massless spin -1 fields. This problem 
was circumvented in \cite{bmss} by assuming non-minimal couplings in the gauge field
sector. An alternative approach was adopted in \cite{pmss} where 
electromagnetism was incorporated in a gauge-invariant fashion into the 
EC framework with the help of a Chern-Simons extension. Such an extension is
natural in the context of a heterotic string theory to ensure anomaly cancellation.
The source of torsion in this case is the two-form Kalb-Ramond (KR) field \cite{kr}
which appears in the massless sector of the heterotic string spectrum \cite{gsw}. 
However, the possibility of parity violation was not addressed in \cite{pmss}.

In this note, we generalize upon the earlier works in several ways. First, the exact
source of the torsion field is explored. The different couplings, especially
those giving rise to parity violation, are obtained in terms of the 
KR fields. 
Secondly, the  Chern-Simons term is utilized to obtain  gauge-invariant
interaction of the torsion field with electromagntism. It is observed that the
Chern-Simons term itself can lead to new parity-violating couplings, although they
are relatively suppressed. Finally, we comment on the existence (or otherwise)
of parity violation when other forms of the pseudo-tensorial extension
are used. Thus we not only achieve a significant generalization of 
the previously obtained results, but also are able to pinpoint the characteristics which
the torsion field must have in order to be a source of parity violation.

The gauge-invariant action for the EC-Maxwell-KR-fermion scenario can be written as

\be
S = \int~ d^{4}x \sqrt{-g} ~\left[~\frac{\tilde{R} (g,{\cal T})}{\kappa} ~-~
    \frac{1}{4} F_{\mu \nu} F^{\mu \nu} ~-~
    \frac{1}{2} {\cal H}_{\mu \nu \lambda} {\cal H}^{\mu \nu \lambda} ~+~
    \frac{1}{\sqrt{\kappa}} {\cal T}^{\mu \nu \lambda} {\cal H}_{\mu \nu 
    \lambda}  ~+~ {\cal L}_{fermion} ~\right]
\ee

\noindent
where ~${\cal T}_{\mu \nu \lambda}$~ and ~${\cal H}_{\mu \nu \lambda}$~ are respectively 
the generalized torsion and KR field strength involving the parity-violating 
extensions 

\bea
{\cal T}^{\mu}_{\nu \lambda} ~&=&~  T^{\mu}_{\nu \lambda} ~+~ q~\left( \epsilon^{\alpha
\beta}_{\nu \lambda}~ T^{\mu}_{\alpha \beta} ~+~ \epsilon^{\mu \sigma}_{\rho [\nu}~
T^{\rho}_{\lambda] \sigma} \right)\\ 
{\cal H}^{\mu}_{\nu \lambda} ~&=&~  \tilde{H}^{\mu}_{\nu \lambda} ~+~ q~\left( \epsilon^{\alpha
\beta}_{\nu \lambda}~ \tilde{H}^{\mu}_{\alpha \beta} ~+~ \epsilon^{\mu \sigma}_{\rho [\nu}~
\tilde{H}^{\rho}_{\lambda] \sigma} \right) 
\eea

Once the covariant derivative $\tilde{D}_{\mu}$~ is defined in terms 
of ~$\tilde{\Gamma}^{\alpha}_{\mu \lambda} ~=~ \Gamma^{\alpha}_{\mu \lambda} ~-~ 
{\cal T}^{\alpha}_{\mu \lambda}$ (where ~$\Gamma^{\alpha}_{\mu \lambda}$~  is the usual 
Christoffel connection), the metricity condition 
~$\tilde{D}_{\mu}~g^{\mu \nu} ~=~ 0$~ is automatically preserved \cite{bmss}.

$\tilde{H}_{\mu \nu \lambda}$, the modified 
KR field strength, can be expressed with a Chern-Simon extension as

\be
\tilde{H}_{\mu \nu \lambda} ~=~ H_{\mu \nu \lambda} ~+~ \sqrt{\kappa}~ 
A_{[\mu} F_{\nu \lambda]}
\ee

It remains invariant under the $U(1)$ gauge transformation ~$\delta A_{\mu} ~=~ \partial_
{\mu} \omega$~ if the KR potential transforms
as ~$\delta B_{\mu \nu} ~=~ - \omega F_{\mu \nu}$. Here ~$ \kappa ~=~ 16 \pi G$~ is the 
gravitational coupling constant and the electromagnetic field tensor 
~$F_{\mu \nu}$~ is 
assumed to be invariant under KR gauge transformation. 
$q$~ is a parameter determining the degree of parity-violation and it presumably depends
on the matter distribution. 
$\tilde{R}(g,{\cal T})$~ is the scalar curvature of the spacetime with torsion, 
defined by ~ $\tilde{R} = \tilde{R}_{\alpha\mu\beta\nu}g^{\alpha\beta}g^{\mu\nu}$.~ 
$\tilde{R}_{\alpha\mu\beta\nu}$~ is the Riemann-Christoffel tensor:

\be
\tilde{R}^{\rho}_{\mu\nu\lambda} ~=~
\partial_{\mu} \tilde{\Gamma}^{\rho}_{{\nu\lambda}}
~-~ \partial_{\nu} \tilde{\Gamma}^{\rho}_{{\mu\lambda}} ~+~ 
\tilde{\Gamma}^{\rho}_{\mu\sigma} \tilde{\Gamma}^{\sigma}_{\nu\lambda} ~-~ 
\tilde{\Gamma}^{\rho}_{\nu\sigma} \tilde{\Gamma}^{\sigma}_{\mu\lambda}  
\ee

\noindent
$\tilde{R}(g,{\cal T})$~ is related to the Einstein scalar curvature ~$R$~ as

\be
\tilde{R}(g,{\cal T}) ~=~ R(g) ~-~ {\cal T}_{\mu \nu \lambda} {\cal T}^{\mu \nu \lambda}
\ee

\noindent
${\cal L}_{fermion}$~ being the Lagrangian density for a Dirac fermion.

The role of the augmented KR field strength three-tensor as the spin angular momentum 
density (which is the source of torsion \cite{hehl}) is evident from Eq.(1) 
where the torsion tensor ${\cal T}_{\mu \nu \lambda}$, being an auxiliary field, 
obeys the constraint equation

\be
{\cal T}_{\mu \nu \lambda} ~=~ \sqrt{\kappa}~ {\cal H}_{\mu \nu \lambda}
\ee

\noindent
and the same relation holds between ~$T$~ and ~$\tilde{H}$~ as well. 

The action for the gravity sector is 

\be
S_{G} ~=~ \int~ d^{4} x ~\sqrt{-g} ~\tilde{R}
\ee

\noindent
where ~$\tilde{R}$~ includes  the contributions due to torsion, and is given by

\be
\tilde{R}(g,{\cal H}) ~=~ R(g) ~-~ \kappa~\tilde{H}_{\mu \nu \lambda} 
\tilde{H}^{\mu \nu \lambda} ~-~ 6 \kappa ~q ~\epsilon_{\alpha \beta}^{\nu \lambda}~
\tilde{H}^{\mu}_{\nu \lambda} \tilde{H}_{\mu}^{\alpha \beta}
\ee

The second term corresponds to the Einstein-Cartan extension of the general theory
of relativity, which conserves parity. The last term which is parity-violating
is identical with that given in \cite{bmss}.
In terms of the KR field strength and the Maxwell field, the parity-violating 
term in the above equation can be expressed as

\be
R^{pv} ~=~  6 \kappa ~q ~\left[ \epsilon_{\alpha \beta}^{\nu \lambda}~
H_{\mu \nu \lambda} ~\left (H^{\mu \alpha \beta} ~+~ 2 \sqrt{\kappa}~ [A^{\mu}
F^{\alpha \beta} ~+~ 2~ A^{\alpha} F^{\beta \mu} ] \right) ~+~ \kappa~ (A_{\mu}
F_{\nu \lambda} ~+~ 4~ A_{\lambda} F_{\mu \nu} )~A^{\mu}~^{*}F^{\nu \lambda} \right]
\ee 

\noindent
where ~$^{*}F^{\nu \lambda} ~=~ \epsilon^{\nu \lambda \alpha \beta}~F_{\alpha
\beta}$~ is the ({\it Hodge}-) dual of $F_{\alpha \beta}$.
The field equations that can be obtained on extremizing the action (8) with respect 
to ~$g_{\mu \nu}$~ are the usual Einstein's equations in the present picture:

\be
R_{\mu \nu} ~-~ \frac 1 2 g_{\mu \nu} R ~=~ \kappa~ \tau_{\mu \nu}
\ee

\noindent
where the symmetric two-tensor ~$\tau_{\mu \nu}$~ has direct analogy with the
energy-momentum tensor for matter couplings and is a clear manifestation of 
spin-density in an extended Einstein-Cartan spacetime. It has the general form 
\cite{ssss}

\be
\tau_{\mu \nu} ~=~ \left( 3 g_{\nu \rho} {\cal H}_{\alpha \beta \mu}
{\cal H}^{\alpha \beta \rho} - \frac{1}{2} g_{\mu \nu} {\cal H}_{\alpha 
\beta \gamma} {\cal H}^{\alpha \beta \gamma} \right)
\ee

The EC-KR theory can be made to couple with electromagnetism   
in a gauge invariant manner in the following way \cite{pmss}:

\be
{\cal L}_{G-EM} ~=~\frac{R}{\kappa} ~-~
    \frac{1}{4} F_{\mu \nu} F^{\mu \nu}  ~-~
    \frac{1}{2} {\cal H}_{\mu \nu \lambda} {\cal H}^{\mu \nu \lambda} 
\ee

\noindent
whence the modified Maxwell equations can be obtained as 

\be
D_{\mu} F^{\mu \nu} ~=~ \sqrt{\kappa}~ {\cal H}^{\mu \nu \lambda} F_{\lambda \nu}
\ee

\noindent
In addition to the above equations, the Maxwell-Bianchi identity 

\be 
D_{\mu}~ ^{*}F^{\mu \nu} ~=~ \frac 1 {\sqrt{- g}} ~\partial_{\mu} ~(\sqrt{- g}~~
^{*}F^{\mu \nu}) ~=~ 0
\ee
 
\noindent
also holds for the electromagnetic field. The KR-Maxwell interaction term in
the action is given by

\be
S_{int} ~=~ \int~ d^{4}x~ \sqrt{- g}~ \left[ H_{\mu \nu \lambda}~ \left( 3~ A^{\mu}~
F^{\nu \lambda} ~+~ 6 q ~ A^{\mu}~^{*}F^{\nu \lambda} ~+~ 12 q~ \epsilon_{\alpha \beta}
^{\nu \lambda}~ A^{\alpha}~ F^{\beta \mu} \right) \right]
\ee

The second and third terms in the expression are the exclusive parity-violating
modifications over the result found in \cite{pmss}. 
If one further expresses the
KR field strength as the ({\it Hodge}-) dual of
the derivative of a pseudoscalar field,  the only non-vanishing parity-violating
contributions can come from the Chern-Simons extension which, however, are
suppressed by the square of the Planck mass. 

For the spin $1/2$~ fermion, the extended Dirac Lagrangian
has the form \cite{aud,fst}

\be
{\cal L}_{fermion} ~=~ \bar{\psi} ~\left[ i \gamma^{\mu}~ \left( \partial_{\mu}
~-~ \sigma^{\rho \beta} v^{\nu}_{\rho} g_{\lambda \nu} \partial_{\mu} v^{\lambda}_
{\beta} ~-~ g_{\alpha \delta} \sigma^{a b} v^{\beta}_{a} v^{\delta}_{b} 
\tilde{\Gamma}^{\alpha}_{\mu \beta} \right) \right] \psi
\ee

\noindent
where the tetrad ~$v^{\lambda}_{b}$~ connects the curved space (designated by Greek
indices) to the corresponding tangent space (designated by Latin indices) at any point.
As shown in \cite{bmss}, using the full form of ~$\tilde{\Gamma}$, ~${\cal L}_{fermion}$~
can be written

\be
{\cal L}_{fermion} ~=~ {\cal L}^{E}_{f} ~+~ {\cal L}^{C}_{f} ~+~ {\cal L}^{pv}_{f} 
\ee

\noindent
where ~${\cal L}^{E}_{f}$~ is the Dirac Lagrangian in 
Einstein gravity, ~${\cal L}^{C}_{f}$~ is its Cartan extension which is  
parity-conserving and can be expressed in terms of the KR field and the Maxwell
field (following the identification (7)) as 

\be
{\cal L}^{C}_{f} ~=~ \bar{\psi}~ \left[ i \gamma^{\mu} \sigma^{a b} v^{\beta}_{a}
v^{\omega}_{b} \left( \sqrt{\kappa}~ H_{\beta \omega \mu} ~+~ \kappa~ (A_{\mu}
F_{\beta \omega} ~+~ 2~ A_{\beta} F_{\omega \mu}) \right) \right] ~\psi
\ee

The additional term which is responsible for parity violation and is denoted 
here by~${\cal L}^{pv}_{f}$, 
has the form

\bea
{\cal L}^{pv}_{f} ~&=&~ q \sqrt{\kappa}~ \bar{\psi}~ \left[ i \gamma^{\mu} 
\sigma^{a b} v^{\beta}_{a}
v^{\omega}_{b} \left( \epsilon_{\mu \beta}^{\gamma \delta}~
H_{\omega \gamma \delta} ~+~ \epsilon^{\gamma \delta}_{\omega [\mu}~ H_{\beta]
\gamma \delta} \right) \right] ~\psi \nonumber \\ 
&+&~ q \kappa~ \bar{\psi}~ \left[ i \gamma^{\mu} 
y\sigma^{a b} v^{\beta}_{a}
v^{\omega}_{b} \left(2~A_{\omega}~ ^{*}F_{\mu \beta} ~-~ 
A_{\mu}~ ^{*}F_{\omega \beta} ~+~ 4~ \epsilon_{\mu \beta}^{\gamma \delta} 
A_{\gamma} F_{\delta \omega}
~-~ 2~ \epsilon_{\omega \beta}^{\gamma \delta} A_{\gamma} F_{\delta \mu} \right)
\right]~ \psi
\eea

Finally, it is worth mentioning here that 
the metricity condition imposes no compelling need to take the generalized
torsion tensor ~${\cal T}^{\mu}_{\nu \lambda}$~ to be antisymmetric in all its
three indices; an antisymmetry in only a pair of indices will suffice. As such,
~${\cal T}^{\mu}_{\nu \lambda}$~ can be expressed in a more general way:

\be
{\cal T}^{\mu}_{\nu \lambda} ~=~  T^{\mu}_{\nu \lambda} ~+~  q_{1}~\epsilon^{\alpha
\beta}_{\nu \lambda}~ T^{\mu}_{\alpha \beta} ~+~ q_{2}~\epsilon^{\mu \sigma}_{\rho 
[\nu}~T^{\rho}_{\lambda] \sigma} 
\ee

\noindent
where the constants ~$q_{1}$~ and ~$q_{2}$ are in general different, and antisymmetry 
is confined to the two lower indices (although ~$T^{\mu \nu \lambda}$~  
is still antisymmetric in all three indices). In the gravity sector, we
then have a Lagrangian density given by

\be
{\cal L}_{gravity} ~=~ \tilde{R} (g, {\cal T}) ~=~ R (g) ~-~ \partial^{\lambda}
{\cal T}^{\alpha}_{\alpha \lambda} ~+~ g^{\mu \lambda}~ \Gamma^{\rho}_{\mu \lambda}~
{\cal T}^{\alpha}_{\alpha \rho} ~+~ {\cal T}^{\alpha \lambda}_{\rho}~
{\cal T}^{\rho}_{\alpha \lambda}
\ee

\noindent
which, in terms of ~$T$,~ can be expressed as

\be
{\cal L}_{gravity} ~=~ \tilde{R} (g, {\cal T}) ~=~ R (g) ~+~ R^{pc} (T) ~+~ 
R^{pv} (\epsilon,T)
\ee
 
\noindent
with

\bea
R^{pc} (T) ~&=&~ - ~[ 1 ~+~ 2 (q_{1} ~+~ 2 q_{2}) ]~ T^{\mu}_{\nu \lambda}~ T_{\mu}^{\nu 
\lambda} \\
R^{pv} (\epsilon,T) ~&=&~ - ~2 (q_{1} ~+~ 2 q_{2})~ \epsilon^{\rho \alpha}_{\beta \sigma}~
T_{\rho \alpha}^{\lambda}~ T_{\lambda}^{\beta \sigma} ~-~ ( q_{1} ~-~ q_{2} )~
\left[ \partial^{\lambda} \left( \epsilon^{\alpha \beta}_{\sigma \lambda}~ T^{\sigma}_
{\alpha \beta} \right) ~-~ g^{\mu \lambda}~ \Gamma^{\rho}_{\mu \lambda}~
\epsilon^{\alpha \beta}_{\sigma \rho}~ T^{\sigma}_{\alpha \beta} \right]
\eea

The presence of the derivatives of ~$T$~ makes it dynamic, i.e., the torsion
tensor ~$T_{\mu \nu \lambda}$~ is no longer an auxiliary field for  ~$q_{1} ~\neq~
q_{2}$. Hence, the identification of ~$T$~ with ~$H$~ is
not possible anymore. The spin - 1/2 fermion field in this case has the Lagrangian
density

\be
{\cal L}_{fermion} ~=~ {\cal L}_{f}^{E} (g) ~+~ {\cal L}_{f}^{pc} (T) ~+~
{\cal L}_{f}^{pv} (\epsilon, T)
\ee

\noindent
where  ${\cal L}_{f}^{E} (g)$ includes the set of terms corresponding to
Einstein gravity, and 

\bea
{\cal L}^{pc}_{f} (T) ~&=&~ \bar{\psi}~ \left[ i \gamma^{\mu} \sigma^{a b} v^{\beta}_{a}
v^{\omega}_{b}~ T_{\mu \beta \omega} \right]~ \psi \\
{\cal L}_{f}^{pv} (\epsilon, T) ~&=&~ q_{1}~ \bar{\psi} ~\left( i \gamma^{\mu} \sigma^{a b} 
v^{\beta}_{a} v^{\omega}_{b}~ \epsilon^{\alpha \delta}_{\mu \beta}~ T_{\alpha \delta \mu}
\right) ~\psi ~+~ q_{2}~ \bar{\psi} ~\left( i \gamma^{\mu} \sigma^{a b} 
v^{\beta}_{a} v^{\omega}_{b}~ \epsilon^{\lambda \nu}_{\omega [\mu}~ T_{\beta] 
\lambda \nu} \right) ~\psi 
\eea

\noindent
which explicitly violates parity.  

One interesting consequence of the above is worth noting here.
As has been already pointed out, there is an alternative way of 
expressing the tensor field $H$
in terms of a pseudoscalar field ~$\phi$~ through a duality transformation:

\be
H^{\mu\nu\lambda} ~=~ \epsilon^{\mu\nu\lambda\alpha}\partial_{\alpha} \phi
\ee

In such a case, the pseudo-tensorial extension of torsion vanishes 
as soon as one sets ~$q_1~=~q_2$;~ in other words, there is no parity violation 
under this condition unless one uses a Chern-Simons extension. On the contrary, 
the inequality of the two charges
retains parity-violating terms in all sectors of the Lagrangian
even when torsion is expressed in terms of the dual field.

We conclude by observing that predictions have already been made 
linking a space-time with torsion with phenomena such as the rotation of
the plane of polarization in radio waves coming from distant galactic
sources \cite{skpm}. Moreover, couplings arising from torsion can affect the helicity
flip of neutrinos \cite{ssas,hfref}, which has far-reaching implications in the context
of the  solar and atmospheric neutrino anomalies. Further insight   
on parity violation due to torsion will help us in evolving better understanding of 
these phenomena.

\bigskip 

{\bf Acknowledgements:} BM and SSG acknowledge partial 
support from Board of Research in Nuclear Sciences,
Government of India, under grant nos. 2000/37/10/BRNS and
98/37/16/BRNS cell/676. The research of SS is funded 
by Council of Scientific and Industrial Research, Government of 
India.

\end{document}